\documentclass[12pt]{article}
\usepackage[margin=1in]{geometry}
\usepackage[utf8]{inputenc}
\usepackage{fullpage, graphicx, multirow, setspace}
\usepackage[svgnames]{xcolor}
\usepackage{amsmath, amsfonts, amssymb}
\usepackage[round]{natbib}
\usepackage{rotating}
\usepackage[section]{placeins}
\usepackage{authblk}
\newcommand{\E}[1]{{\mbox{E}}\left[#1\right]}
\newcommand{\Var}[1]{{\mbox{Var}}\left(#1\right)}

\title{Evaluating Approximations of Count Distributions and Forecasts for Poisson-Lindley Integer Autoregressive Processes}
\author[1]{Rachel D.~Gidaro\footnote{Department of Mathematical Sciences, United States Military Academy, West Point}}
\author[2]{Jane L.~Harvill}
\affil[1]{e-mail: rachel\_gidaro1@baylor.edu \\ Department of Statistical Science, Baylor University, Waco, TX 76798-7140}
\affil[2]{e-mail: jane\_harvill@baylor.edu \\ Department of Statistical Science, Baylor University, Waco, TX 76798-7140
  \\ ORCID ID: \#0000-0001-8585-5945}
\date{June 2024 \\ Word Count: 3861}

\begin{document}

\maketitle
\newpage
\begin{abstract}
Although many time series are realizations from discrete processes, it is often that a continuous Gaussian model is implemented for modeling and forecasting the data, resulting in incoherent forecasts.  Forecasts using a Poisson-Lindley integer autoregressive (PLINAR) model are compared to variations of Gaussian forecasts via simulation by equating relevant moments of the marginals of the PLINAR to the Gaussian AR. To illustrate utility, the methods discussed are applied and compared using a discrete series with model parameters being estimated using each of conditional least squares, Yule-Walker, and maximum likelihood.

\noindent
{\bf{Keywords:}} Discrete time series, coherent forecasts, Poisson-Lindley integer autoregressive process, Kullback-Leibler divergence, Kolmogorov metric.
\end{abstract}

\section{Introduction}
\label{sec:intro}
Time series analysis is used across a variety of areas of scientific inquiry, from business to health sciences, physics, geology, and social sciences, to name a few. In many circumstances, researchers look to model a process over time and ultimately forecast future events. Traditional time series analysis uses  
methods based on continuous processes, such as an autoregressive-moving average (ARMA) with Gaussian innovations, for analysis and forecasting, even when the process is discrete. In fact many time series textbooks commonly do this, see \cite{boxjen76}, \cite{Shumway_Stoffer_2017}, and \cite{Brockwell_Davis_1996}.

One desirable property of a time series model is that it produces forecasts that do not violate known constraints; that is, the model produces ``coherent forecasts.'' Traditional application of a continuous ARMA model for forecasting a discrete process will result in violating coherency.  Thus, coherent forecasting is one area where discrete modeling can yield improvements over the traditional continuous models.  In a series of seminal papers~\citep{Jacobs_Lewis_1978a, Jacobs_Lewis_1978b, Jacobs_Lewis1983}, Jacobs and Lewis introduced the theory and analysis for discrete time series processes.  The disadvantages of these early models are well-documented; see, for example~\cite{aanda1987, aanda1988}.  

A collection of discrete time series models that addresses some of the disadvantages of Jacobs and Lewis incorporates binomial thinning.  The first of these applies binomial thinning to a process with Poisson marginals, yielding the first-order integer-valued autoregressive (INAR(1)) model~\citep{aanda1987, mckenzie1988}. To address overdispersion, \cite{aanda1993} developed an INAR(1) process with generalized Poisson marginals. Other approaches to addressing overdispersion apply binomial thinning to processes with geometric or negative binomial marginals~\citep{mckenzie1986}.
The Poisson-Lindley distribution was introduced by \cite{Sankaran_1970} to model count data.  This distribution is unimodal, has larger variability than the traditional Poisson distribution, and has smaller skewness and kurtosis than the negative binomial model \citep{ghitany2009}. In the time series framework, \cite{moham2018} introduced the INAR(1) process with Poisson-Lindley marginals. One downside to the Poisson-Lindley INAR(1) model is that it can be complex and difficult to estimate the model parameters. 

Many papers have looked at addressing the efficacy of Gaussian methods applied to discrete time series. In \cite{mandb2016}, various INAR(1) processes were compared to the Gaussian autoregressive process, where the forecast for the Gaussian model was rounded to the nearest integer.  They found that the Gaussian autoregressive process produced similar results to that of the INAR(1) process. However, in \cite{aandw2016} the marginal forecast distribution of INAR(1) processes was approximated by the Gaussian AR(1) conditional forecast distribution, and it was found that the approximation was worse, especially in the range of low counts data.  Additionally \cite{bisandgero2015} showed when systematically comparing a coherent point forecast method of an INAR(1) model to the approximate point forecast obtained from a Gaussian AR(1), the Gaussian had higher forecast mean square error and forecast mean absolute error.  \cite{homburg2020} thoroughly investigated the differences between the Poisson INAR(1) to the Gaussian AR(1) using a variety of measures
to compare the forecasts from the competing models. Her main conclusion is the quality of the approximation is influenced by the amount of autocorrelation present.  Finally, ~\cite{gidaroandharvill2024} provide a thorough investigation of forecasting discrete time series using a GINAR(1).  In their work, they compare the conditional distributions for a discretized Gaussian AR(1) from two different moment-equated methods to the GINAR(1) conditional distribution.  They also compare the forecasts of the different models that takes into account different parameter estimation techniques.  They find that in some circumstances, the discretized Gaussian forecasts performs as well as the forecasts based on the GINAR model, but in most cases, the GINAR outperforms the Gaussian AR forecasts.

In what follows, we investigate the performance of Poisson-Lindley forecasts compared to their Gaussian counterpart. The rest of the paper is organized as follows.  In Section~\ref{sec:PLINAR1} the first order Poisson-Lindley integer autoregressive process (PLINAR(1)), first introduced by~\citet{moham2018}, is defined. In Section~\ref{sec:forecast}, we present the $k$-step-ahead conditional mean forecasts for a PLINAR(1) process and for a traditional Gaussian first order autoregressive (AR(1)) model.  In Section~\ref{sec:comparisons}, we develop two methods for equating moments of the PLINAR(1) to the Gaussian AR(1). The closeness of the conditional PMFs for the forecasting methods are compared in Section~\ref{sec:comparisons} via Kullback-Leibler divergence and Kolmogorov metric.  To illustrate utility and to further investigate efficacy, the methods are then applied to a discrete time series data set.  Prior to that, in Section~\ref{sec:appintro}, three methods for parameter estimation are presented along with criteria for comparing the forecasts. Concluding remarks are in Section~\ref{sec:conclusion}.

\section{Poisson-Lindley Integer-Valued Autoregressive Processes}
\label{sec:PLINAR1}
The Poisson-Lindley integer-valued autoregressive model of order one, or PLINAR(1), first proposed by \citet{moham2018} is defined as follows.  Let $B_j$ be independent and identically distributed Bernoulli($\alpha$) random variables where for each $j,~P(B_j = 1) = \alpha$.  The binomial thinning operator $\circ$ is
\begin{equation*}
\label{eq:thinning}
  \alpha \circ X = \sum_{j=1}^X B_j.
\end{equation*}
\citep{steutel1979}.  The process $\{X_t\}$ follows a PLINAR(1) if 
\begin{equation*}
\label{eq:PLINAR1}
  X_t = \alpha \circ X_{t-1} + I_tH_t, \qquad t = 1, 2, \ldots, \quad 0 \le \alpha \le 1,
\end{equation*}
where $I_t$ has $P(I_t = 0) = 1 - P(I_t = 1) = \alpha$, and for all $t,~H_t$ has probability mass function
\begin{align*}
\label{eq:genmix}
\nonumber
    g(h) = P(H_t = h) & = \frac{\theta^2(1 - \alpha)^2 + \theta(1 - \alpha^2) + 2\alpha}{\left\{\theta(1 - \alpha) + 1)^2\right\}} \frac{\theta}{1 + \theta}\left(1 - \frac{\theta}{1 + \theta}\right) \\
\nonumber
    & \qquad + \frac{(1 - \alpha)}{\theta(1 - \alpha) + 1}(h + 1)\left(\frac{\theta}{1 + \theta}\right)^2 \left(1 - \frac{\theta}{1 + \theta}\right)^h \\
\nonumber
    & \qquad - \frac{\alpha}{\left\{\theta(1 - \alpha) + 1\right\}^2}\frac{\theta + 1}{\theta + 1 + \alpha}\left(1 - \frac{\theta + 1}{\theta + 1 + \alpha}\right)^h, \qquad h = 0, 1, 2, \ldots.
\end{align*}
It is further assumed that $X_0$ is a PL($\theta$) random variable. \cite{Sankaran_1970} showed that the probability mass function (PMF) can be written as follows:
\begin{equation*}
    \label{eq:PLPMF}
  P(X_0 = x) = \frac{\theta^2(x + \theta + 2)}{(1 + \theta)^{x + 3}}, \qquad x = 0, 1, 2, \ldots, \quad \theta > 0 .
\end{equation*}
Then, the mean and variance of the Poisson-Lindley($\theta$) distribution are
$ (\theta + 2)/\{\theta(\theta + 1)\}$ and
$(\theta^3 + 4\theta^2 + 6\theta + 2)/\left\{\theta^2(\theta + 1)^2\right\}$,
respectively.  Since the marginal distributions of the $X_t$ are Poisson-Lindley, it follows that \begin{equation*}
\label{eq:meanandvarofx}
\E{X_t} = \E{H_t} = \frac{\theta + 2}{\theta(\theta + 1)}
\quad {\text{and}} \quad 
\Var{X_t} = \frac{\theta^3 + 4\theta^2 + 6\theta + 2}{\theta^2(\theta + 1)^2}.  
\end{equation*}

\section{Forecasting a PLINAR Process}
\label{sec:forecast}
Traditionally, given a series of length $n,~\mathbf{X}_n = (X_n, \ldots, X_1)$, the $k$-step-ahead forecast is the conditional mean $\E{X_{n+k}\,|\,\mathbf{X}_n}$. The $k$-step ahead conditional PMF for the PLINAR(1), given by \cite{moham2018} is
\begin{equation*}
\label{eq:kstepPMF}
    P(X_{n+k} = y\,|\,\mathbf{X}_n, X_n = x) =
    \sum_{j=0}^{\min(y, x)} \binom{x}{j} \alpha^{kj}(1- \alpha^k)^{x-j} P(Z_{n+k} = y - j),
\end{equation*}
where
\begin{equation*}
\label{eq:innovkstepPMF}
    P(Z_{n+k} = z) = 
    \left\{
    \begin{array}{ll}
    \alpha^k + (1 - \alpha^k)\left[A_k\frac{\theta}{1 + \theta} + B_k\left(\frac{\theta}{1 + \theta}\right)^2 + C_k \frac{1 + \theta}{1 + \theta+ \alpha^k}\right]
    & {\mbox{if $z = 0$,}} \\\\
    (1 - \alpha^k)\left[A_k\frac{\theta}{1 + \theta}\left(\frac{1}{1 + \theta}\right)^z + B_k(z + 1)\left(\frac{\theta}{1 + \theta}\right)^2\left(\frac{1}{1 + \theta}\right)^z  + C_k \frac{1 + \theta}{1 + \theta+ \alpha^k}\left(\frac{\alpha^k}{1 + \theta+\alpha^k}\right)^z \right]
    & {\mbox{if $z = 1, 2, \hdots,$}} \\
    \end{array}\right.
\end{equation*}
and
\begin{equation*}
\label{eq:innovkstepPMFpieces}
    A_k = \frac{\theta^2(1 - \alpha^k)^2 + \theta(1 - \alpha^{2k}) + 2\alpha^k}{\left\{\theta(1 - \alpha^k) + 1\right\}^2},\qquad
    B_k = \frac{1 - \alpha^k}{\theta(1 - \alpha^k) + 1}, \qquad
    C_k = \frac{-\alpha^k}{\left\{\theta(1 - \alpha^k) + 1\right\}^2}.
\end{equation*}
The $k$-step ahead conditional mean and variance provided by \cite{moham2018} are, respectively
\begin{align}
\label{eq:condmean}
\E{X_{n+k}\,|\,\mathbf{X}_n, X_n = x} & = \alpha^k x + (1 - \alpha^k)\frac{\theta + 2}{\theta(\theta + 1)} \\
\nonumber
\Var{X_{n+k}\,|\,\mathbf{X}_n, X_n = x} & = \alpha^k(1 - \alpha^k)x + \left(\frac{1 - \alpha^{2k}}{1-\alpha^2}\right) \sigma^2_\epsilon + \left[\frac{(1 - \alpha^k)(\alpha - \alpha^{k + 1})}{1 - \alpha^2}\right]\mu_\epsilon,
\end{align}
where $\mu_\epsilon$ and $\sigma^2_\epsilon$ are the marginal mean and variance of $\epsilon_t$ given by
\begin{equation*}
\label{eq:margmeanvareps}
\mu_\epsilon = \frac{(1 - \alpha)(\theta + 2)}{\theta(\theta + 1)} \quad
{\mbox{and}} \quad
\sigma^2_\epsilon = \frac{(1 - \alpha)\left[\theta^3 + 4\theta^2 + 6\theta + 2 + \alpha(\theta^2 + 4\theta + 2)\right]}{\theta^2(\theta + 1)^2}.
\end{equation*}
As $k \rightarrow \infty, \E{X_{n+k}\,|\,\mathbf{X}_n} \rightarrow \E{X_n}$ and
$\Var{X_{n+k}\,|\,\mathbf{X}_n} \rightarrow \Var{X_n}$. 

As noted earlier, it is common practice to use a Gaussian autoregressive process (AR) for modeling and forecasting discrete-valued time series.  If $\{W_t\}$ denotes a Gaussian autoregressive process of order $p$, AR($p$), $W_t$ can be written
\begin{equation*}
  \label{eq:ARp}
  W_t = \phi_0 + \phi_1 W_{t-1} + \cdots + \phi_p W_{t-p} + \varepsilon_t, \quad t = 1, 2, \ldots,
\end{equation*}
where $\varepsilon_t$ are IID normal random variables with finite mean $\mu_\varepsilon$ and finite variance $\sigma^2_\varepsilon$, independent of the $\{W_s:\, s < t\}$.  If the mean of the innovations $\E{\varepsilon_t} = 0$, then $\phi_0 = 0$.
Assuming the process is second-order stationary, the autocovariance function of $W_t$ satisfies the Yule-Walker equations.  In particular, for the AR(1) with $\phi = \phi_1$, 
the marginal mean and variance of $W_t$ are 
\begin{equation*}
\label{eq:AR1meanandvar}
\E{W_t} = \frac{\mu_\varepsilon}{1 - \phi} \quad {\mbox{and}} \quad
\Var{W_t} = \frac{\sigma^2_\varepsilon}{1 - \phi^2}.
\end{equation*}
Let ${\mathbf{W}}_n = (W_n, \ldots, W_1)$.  Then for $W_n = w$, the $k$-step-ahead forecast is the $k$-step conditional mean given by
$$
\E{W_{n+k}\,|\,{\mathbf{W}}_n, W_n = w} = \phi^k w,
$$
the $k$-step-ahead conditional distribution is normal, 
\begin{equation*}
\label{eq:kstepcondnormal}
W_{n+k}\,|\,\mathbf{W}_n \sim {\mbox{N}}\left(\mu_{W_{n+k}\,|\,\mathbf{W}_n},
\sigma^2_{W_{n+k}\,|\,\mathbf{W}_n}\right),
\end{equation*}
where for $W_n = w$
\begin{align}
\label{eq:gausscondmean}
\mu_{W_{n+k}\,|\,\mathbf{W}_n} & = \phi^k w + \mu_\varepsilon\frac{1 - \phi^k}{1 - \phi} \\
\label{eq:gausscondvar}
\sigma^2_{W_{n+k}\,|\,\mathbf{W}_n} & = \left(\frac{1 - \phi^{2k}}{1 - \phi^2}\right)\sigma^2_\varepsilon,
\end{align}

\section{Comparing PLINAR(1) to Gaussian AR(1) Predictions}
\label{sec:comparisons}

In order to investigate how well the Gaussian AR(1) process can be used for $k$-step-ahead forecasting of a PLINAR(1) process, two approaches were considered for equating first- and second-moments.  The first approach, which we refer to as the ``innovation method,'' equates the means and the variances of the innovation series; the second equates the means and variances of the marginal distributions.  We refer to the second method as the ``marginal method.''  For both, the Gaussian AR(1) coefficient $\phi$ is set equal to the PLINAR(1) thinning parameter $\alpha$.  We note that since $0 < \alpha < 1$, the Gaussian AR(1) will be second-order stationary.

To equate the mean and variance of the innovations, let
\begin{align}
 \label{eq:e-method}
    \nonumber
    \E{\varepsilon} & = \E{\epsilon}  & \Var{\varepsilon} & = \Var{\epsilon} \\
    \mu_\varepsilon & = \frac{(1 - \alpha)(\theta + 2)}{\theta(\theta + 1)}  & \sigma^2_\varepsilon&= \frac{(1 - \alpha)\left[\theta^3 + 4\theta^2 + 6\theta + 2 + \alpha(\theta^2 + 4\theta + 2)\right]}{\theta^2(\theta + 1)^2}.
\end{align}
Thus, for the innovations method, $\mu_\varepsilon$ and $\sigma^2_\varepsilon$ in~\eqref{eq:gausscondmean} and~\eqref{eq:gausscondvar} are replaced by the corresponding expressions in~\eqref{eq:e-method} for computing $k$-step-ahead forecasts using the Gaussian AR(1).

For the marginal method, we first find an expression for $\mu_\varepsilon$ so that the marginal means of $X_t$ and $W_t$ are equal.  This gives
\begin{align}
\nonumber
    \E{W_t} &= \E{X_t} \\
\nonumber    
    \frac{\mu_\varepsilon}{1 - \alpha} & = \frac{\theta + 2}{\theta (\theta + 1)}   \\
\label{eq:x-methodmean}
    \mu_\varepsilon & = \frac{(1 - \alpha)(\theta + 2)}{\theta (\theta + 1)},
\end{align}
which is equal to the expression for $\mu_\varepsilon$ in~\eqref{eq:e-method}.  However, to equate the marginal variances, $\sigma^2_\varepsilon$ is
\begin{align}
\nonumber
  \Var{W_t} &= \Var{X_t} \\
\nonumber  
   \frac{\sigma^2_\varepsilon}{1 - \alpha^2} &= \frac{\theta^3 + 4\theta^2 + 6\theta + 2}{\theta^2(\theta + 1)^2}  \\
\label{eq:x-methodvar}
  \sigma^2_\varepsilon &= \frac{(1 - \alpha^2)(\theta^3 + 4\theta^2 + 6\theta + 2)}{\theta^2(\theta + 1)^2}
\end{align}
These expressions are used in~\eqref{eq:gausscondmean} and~\eqref{eq:gausscondvar} to find $k$-step-ahead forecasts using the Gaussian AR(1).  Note that denominators in the expressions for the variances in~\eqref{eq:e-method} and~\eqref{eq:x-methodvar} are equal.  It can be shown that for all $\theta > 0$, the numerator of the variance for the innovation method is less than the numerator of the variance for the marginal method.  Specifically, as $\alpha$ gets closer to one and as $\theta$ increases the marginal variance becomes larger faster than the innovation variance. Consequently, there is less variability in the forecasts computed when equating innovation moments than in equating marginal moments.  For values considered in our simulation study in Section~\ref{sec:simulation}, the ratio of the variances ranges from 1.069444 when $\alpha = 0.1$ and $\theta = 0.5$, to 1.780168 when $\alpha = 0.9$ and $\theta = 5$.

\section{Assessing Forecasts}
\label{sec:assess}

In this section, we investigate how well the conditional PMF based on the Gaussian AR(1) compares to the conditional PLINAR(1) PMF.   Our investigation is similar to that of~\cite{gidaroandharvill2024} and~\cite{homburg2020}.  \citeauthor{homburg2020} considers a Poisson AR(1) and sets $X_n = $ round($\mu$).  Here, for the Poisson-Lindley AR(1), we consider the effects of a variety of combinations of values of $\alpha, \theta$, and $X_n$.  Additionally, our target distribution is from a PLINAR(1), compared to Poisson. 
 Before presenting the results of the study, a brief overview of the Kullback-Leibler divergence and the Kolmogorov metric is provided.

\subsection{Kullback-Leibler Divergence}
\label{sec:kld}
The Kullback-Leibler divergence (KLD) is a commonly used statistical distance that measures the dissimilarity between two different probability distributions and is defined as
\begin{equation*}
    KL_{p,q} = \sum_y p(y) \log\left(\frac{p(y)}{q(y)}\right)
\end{equation*}
where $p(y)$ is the target distribution function and $q(y)$ is the approximation of $p(y)$.  The Kullback-Leibler divergence is a nonsymmetric measure of the aggregate disagreement of two distribution functions.  Exact agreement will result in $KL_{p,q} = 0$.  Larger values of $KL_{p,q}$ imply the two distributions are more dissimilar.  For the purposes of this study, we note that $p(y)$ is the conditional PMF given in~\eqref{eq:kstepPMF}.  Since the conditional distribution of the Gaussian AR(1) is zero for all $y = 0, 1, \ldots$, an equivalent measure to~\eqref{eq:kstepPMF} must be found for $q(y)$.  To accomplish this, let $\Phi(\cdot)$ represent the standard normal CDF, and define $\tilde G(\cdot)$ as
$$
\tilde G(y) = \Phi\left(\frac{y - \mu}{\sigma}\right),
\quad {\mbox{for all $y = 0, 1, \ldots$,}}
$$
where expressions for the mean $\mu$ and the standard deviation $\sigma = \sqrt{\sigma^2}$ are provided in~\eqref{eq:e-method}, or~\eqref{eq:x-methodmean} and~\eqref{eq:x-methodvar}, depending upon which approximation method is being evaluated.  Thus $q(y)$ is defined as
$$
q(y) = \tilde G(y) - \tilde G(y - 1) \quad
{\text{for $y = 1, 2, \ldots.$}}
$$
At $y = 0,~q(0) = \tilde G(0)$.

\subsection{Kolmogorov Metric}
\label{sec:km}
Although the Kullback-Leibler divergence is an effective method for quantifying the dissimilarity between two different probability measures, it is not a metric; therefore, $KL_{p,q} \neq KL_{q,p}$.  For this reason, we also consider the Kolmogorov metric which is the maximum distance between any two CDFs and is defined as
\begin{equation*}
    K_{F,G} = \max_y\bigl|F(y) - G(y)\bigr|.
\end{equation*}
Interpreting the Kolmogorov metric is straight-forward compared interpreting $KL_{p.q}$.   Simply put, the Kolmogorov metric is the maximum distance between the target cumulative distribution function $F(\cdot)$ from the approximating $G(\cdot)$ across all values of $y$.  In the context of comparing forecasts, $F(\cdot)$ is the conditional cumulative distribution function (CDF) for the PLINAR(1) process based on the conditional PMF in~\eqref{eq:kstepPMF} and $G(\cdot)$ is the conditional CDF for the Gaussian AR(1) process given in~\eqref{eq:kstepcondnormal} for $k = 1$.

\subsection{Simulation Study}
\label{sec:simulation}
\begin{figure}[!h]
   \centering
   \includegraphics[width=\textwidth]{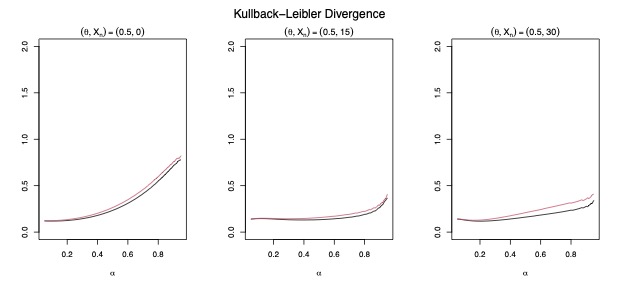}
   \caption{Kullback-Leibler divergences for the marginal method (black) and the innovation method (red) when $\theta = 0.5$.}
   \label{fig:kldtheta1}
 \end{figure}

\begin{figure}[!h]
   \centering
   \includegraphics[width=\textwidth]{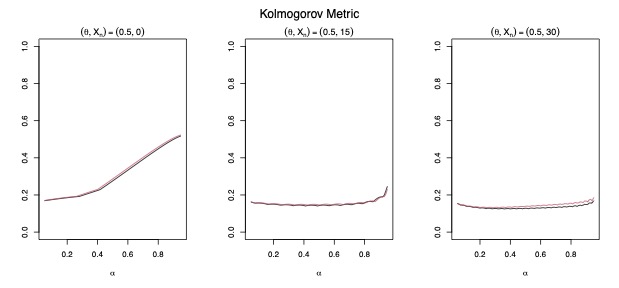}
   \caption{Kolmogorov Metric for the marginal method (black) and the innovation method (red) when $\theta = 0.5$.}
   \label{fig:kolmtheta1}
\end{figure}

\begin{figure}[!h]
   \centering
   \includegraphics[width=\textwidth]{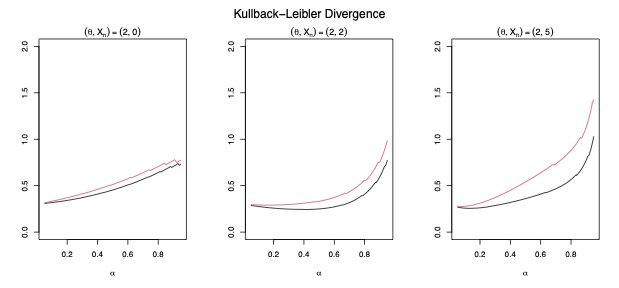}
   \caption{Kullback-Leibler divergences for the marginal method (black) and the innovation method (red) when $\theta = 2$.}
   \label{fig:kldtheta5}
\end{figure}

\begin{figure}[!h]
   \centering
   \includegraphics[width=\textwidth]{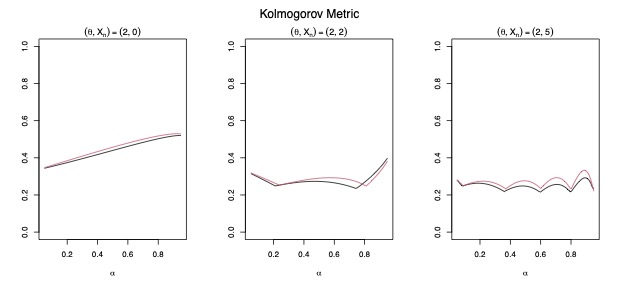}
   \caption{Kolmogorov Metric for the marginal method (black) and the innovation method (red) when $\theta = 2$.}
   \label{fig:kolmtheta5}
\end{figure}

\begin{figure}[!h]
   \centering
   \includegraphics[width=\textwidth]{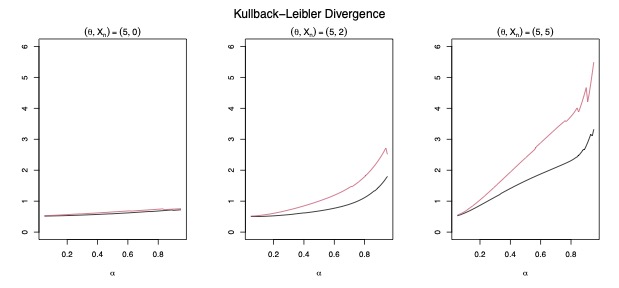}
   \caption{Kullback-Leibler divergences for the marginal method (black) and the innovation method (red) when $\theta = 5$.}
   \label{fig:kldtheta9}
\end{figure}

\begin{figure}[!h]
   \centering
   \includegraphics[width=\textwidth]{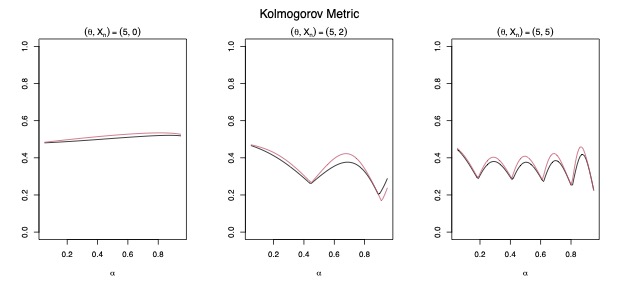}
   \caption{Kolmogorov Metric for the marginal method (black) and the innovation method (red) when $\theta = 5$.}
   \label{fig:kolmtheta9}
\end{figure}

\begin{figure}[!h]
   \centering
   \includegraphics[width=\textwidth]{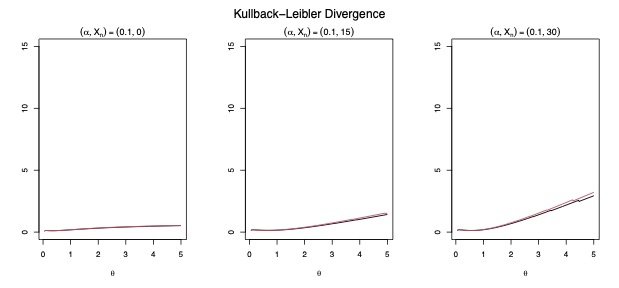}
   \caption{Kullback-Leibler divergences for the marginal method (black) and the innovation method (red) when $\alpha = 0.1$.}
   \label{fig:kldalpha1}
\end{figure}

\begin{figure}[!h]
   \centering
   \includegraphics[width=\textwidth]{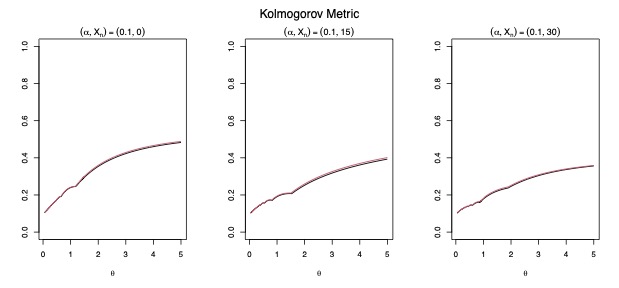}
   \caption{Kolmogorov Metric for the marginal method (black) and the innovation method (red) when $\alpha = 0.1$.}
   \label{fig:kolmalpha1}
\end{figure}

\begin{figure}[!h]
   \centering
   \includegraphics[width=\textwidth]{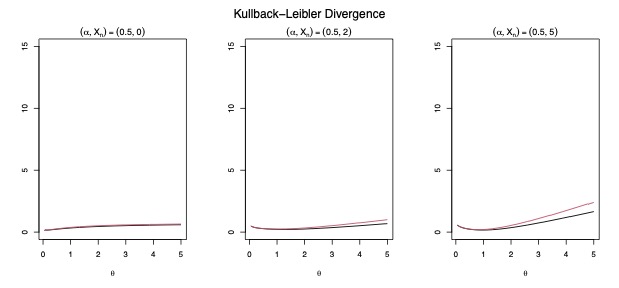}
   \caption{Kullback-Leibler divergences for the marginal method (black) and the innovation method (red) when $\alpha = 0.5$.}
   \label{fig:kldalpha5}
\end{figure}

\begin{figure}[!h]
   \centering
   \includegraphics[width=\textwidth]{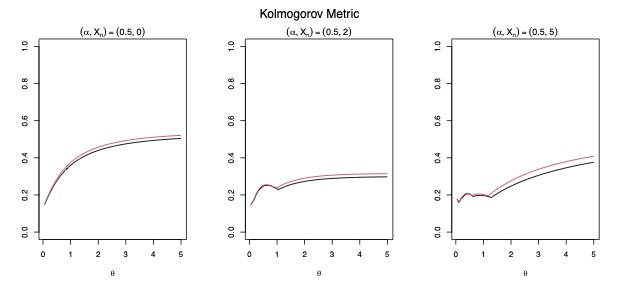}
   \caption{Kolmogorov Metric for the marginal method (black) and the innovation method (red) when $\alpha = 0.5$.}
   \label{fig:kolmalpha5}
\end{figure}

\begin{figure}[!h]
   \centering
   \includegraphics[width=\textwidth]{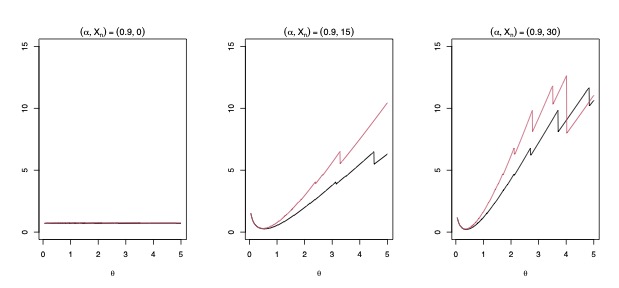}
   \caption{Kullback-Leibler divergences for the marginal method (black) and the innovation method (red) when $\alpha = 0.9$.}
   \label{fig:kldalpha9}
\end{figure}

\begin{figure}[!h]
   \centering
   \includegraphics[width=\textwidth]{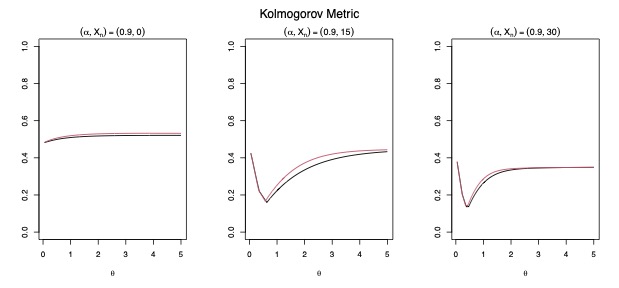}
   \caption{Kolmogorov Metric for the marginal method (black) and the innovation method (red) when $\alpha = 0.9$.}
   \label{fig:kolmalpha9}
\end{figure}

\begin{figure}[!h]
   \centering
   \includegraphics[width=\textwidth]{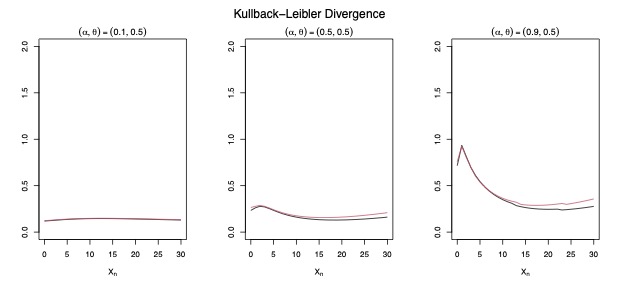}
   \caption{Kullback-Leibler divergences for the marginal method (black) and the innovation method (red) when $\theta = 0.5$.}
   \label{fig:kldxntheta1}
\end{figure}

\begin{figure}[!h]
   \centering
   \includegraphics[width=\textwidth]{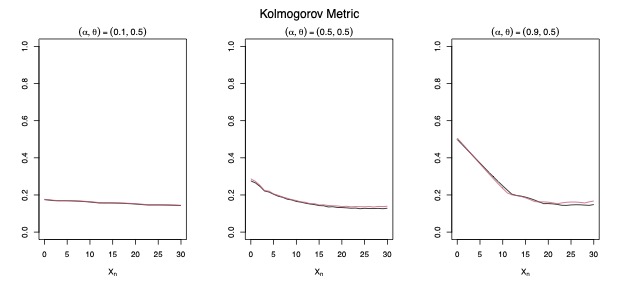}
   \caption{Kolmogorov Metric for the marginal method (black) and the innovation method (red) when $\theta = 0.5$.}
   \label{fig:kolmxntheta1}
\end{figure}

\begin{figure}[!h]
   \centering
   \includegraphics[width=\textwidth]{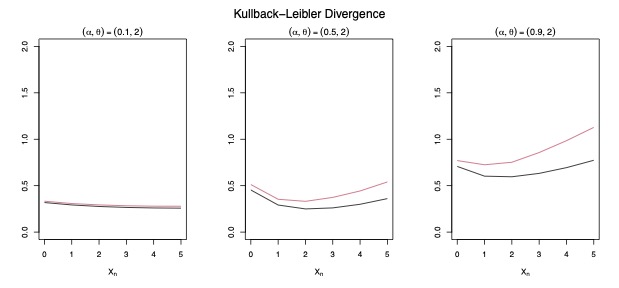}
   \caption{Kullback-Leibler divergences for the marginal method (black) and the innovation method (red) when $\theta = 2$.}
   \label{fig:kldxntheta5}
\end{figure}

\begin{figure}[!h]
   \centering
   \includegraphics[width=\textwidth]{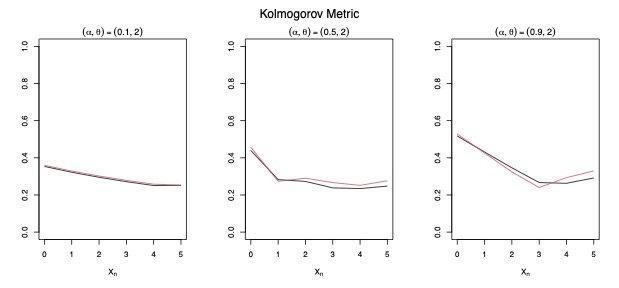}
   \caption{Kolmogorov Metric for the marginal method (black) and the innovation method (red) when $\theta = 2$.}
   \label{fig:kolmxntheta5}
\end{figure}

\begin{figure}[!h]
   \centering
   \includegraphics[width=\textwidth]{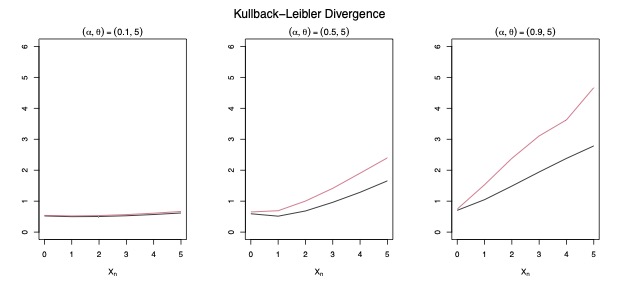}
   \caption{Kullback-Leibler divergences for the marginal method (black) and the innovation method (red) when $\theta = 5$.}
   \label{fig:kldxntheta9}
\end{figure}

\begin{figure}[!h]
   \centering
   \includegraphics[width=\textwidth]{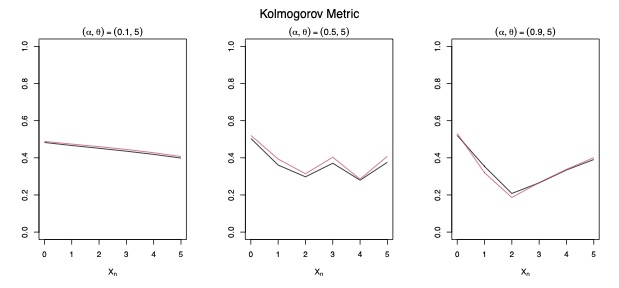}
   \caption{Kolmogorov Metric for the marginal method (black) and the innovation method (red) when $\theta = 5$.}
   \label{fig:kolmxntheta9}
\end{figure}

A simulation study was conducted to compare the performances of one-step-ahead ($k = 1$) forecasting using a PLINAR(1) model versus a Gaussian AR(1).  Comparisons were based on $KL_{p.q}$ and $K_{F,G}$ for all combinations of $\alpha = 0.1, 0.5$ and 0.9, and $\theta = 0.5, 2$, and 5.  Results are visually displayed in Figures~\ref{fig:kldtheta1} through~\ref{fig:kolmxntheta9}.  The figures are organized as follows.  Each figure contains three graphs that show the Kullback-Leibler divergence or the Kolmogorov metric for combinations of values of $\alpha, \theta$, and $X_n$.  On each graph, are two curves.  The black curve is the $KL_{p,q}$ or $K_{F,G}$ for the marginal method; the red is for the innovation method. 

Figures~\ref{fig:kldtheta1} and~\ref{fig:kolmtheta1} contain the two distances for $\theta = 0.5$, $X_n = 0, 15$, and 30, and $\alpha$ ranging from 0.05 to 0.95 by increments of 0.01.  For $\theta = 0.5$, we note that $\E{X_n} = 3.3333$, which is why we consider larger values for $X_n$.  We notice that the marginal and innovation methods perform equally well. In this case, because the prediction variance of the innovation method is smaller than the prediction variance of the marginal method, the innovation method would be preferred. 

Figures~\ref{fig:kldtheta5} and~\ref{fig:kolmtheta5} provide similar information for $\theta = 2$ and $X_n = 0, 2$, and 5, and  Figures~\ref{fig:kldtheta9} and~\ref{fig:kolmtheta9} for $\theta = 5$ and $X_n = 0, 2$, and 5.  Maximum values for $X_n$ were chosen based on the mean $\E{X_n}$.  Overall, the conclusions are the same.  However, we note that values of the Kullback-Leibler divergences are larger as $\theta$ and $\alpha$ increase. So the marginal prediction method may be preferred in these cases even though it has a larger variance.  

A prominent feature of the graphs containing the Kolmogorov distance is the increasing and decreasing pattern.  This pattern is governed by values of $\theta, \alpha$ and $X_n$.  Specifically,~\eqref{eq:gausscondmean} shows the conditional mean of the forecast using the Gaussian approximation includes an additive term $\mu_\varepsilon = (1 - \alpha)(\theta + 2)/(\theta (\theta + 1))$.  Since $k = 1$, the factor $(1 - \phi^k)/(1 - \phi) = 1$.  When $\theta$ is larger, the local minimums of the Kolmogorov distances occur close to integer values of $\alpha X_n$. However, as $\theta$ becomes smaller, $\mu_\varepsilon$ becomes more heavily weighted in the conditional mean. 

Figures~\ref{fig:kldalpha1} through~\ref{fig:kolmalpha9} contain the Kullback-Leibler and Kolmogorov values for values of $\alpha = 0.1, 0.5$, and 0.9 and values of $X_n$ determined by $\E{X_n}$ as for the previous study.  However, in the following figures the value of $\theta$ ranges from 0.05 to 5 in increments of 0.01.  When comparing the distances via $KL_{p,q}$ and $K_{F,G}$, the innovation method is still outperformed by the marginal method for one-step-ahead forecasting.  In most cases, the improvement in these distances is slight or negligible, and so since the innovation method has the smaller forecast error, one could argue in favor of either method.  

In the graphs of the Kullback-Leibler divergence for $\alpha = 0.9$ and $X_n = 15$ or 30, an irregularity occurs.  This can be explained by noting that for $\alpha = 0.9$, the value of $X_{n+1}$ would be expected to be similar to that of $X_n$.  However, as $\theta$ increases the mean would become smaller, which contraindicates many values of $X_n$ being as large as 15.  The erratic behavior is even more pronounced when $X_n = 30$.  In our applications, in Section~\ref{sec:applications}, the largest estimated $\alpha$ is 0.3942.  Therefore these irregularities are noteworthy, but perhaps not too important in determining which prediction method should be used for those data sets. 

Finally, Figures~\ref{fig:kldxntheta1} through~\ref{fig:kolmxntheta9} show $KL_{p,q}$ and $K_{F,G}$ for fixed values of $\alpha = 0.1, 0.5$, and 0.9, and $\theta = 0.5, 2$, and 5, allowing $X_n = 0, 1, 2, \ldots$, up to 5 or 30, depending upon $\E{X_n}$ as previously described.  When $\alpha = 0.1$ the performances of the two methods are essentially the same as measured by the two distances.  Therefore the innovation method might be preferred as it has a smaller one-step-ahead prediction error.  When $\alpha = 0.5$ and $\theta = 2$ or 5, the marginal method has smaller distances, and so would be the preferred method.  When $\alpha = 0.5$ and $\theta = 0.5$ the two are close to equivalent.  Lastly, when $\alpha = 0.9$ the marginal method outperforms the innovation method expect when $\theta = 0.5$, when the two are equivalent.  In summary, for combinations of small to moderate $\alpha$ and small $\theta$, the marginal method appears to be more accurate in one-step-ahead prediction.


\FloatBarrier

\newpage

\section{Estimation of Parameters for the PLINAR(1) Model}
\label{sec:appintro}

In what follows, we will use the forecasting algorithms described in Section~\ref{sec:forecast} to find $k$-step ahead forecasts for a discrete time series.  To accomplish this, model parameters must be estimated.  Therefore in our application section, we investigate the effect of three different estimation techniques, described in Section~\ref{sec:estimation}, on the forecasts using three different forecast measures provided in Section~\ref{sec:accuracy}.

\subsection{Estimation Techniques}
\label{sec:estimation}
Three techniques are prevalent in the literature for estimating the parameters of the PLINAR(1) model defined in Section~\ref{sec:PLINAR1}; Yule-Walker, conditional least squares, and maximum likelihood.  We provide a brief overview of the three techniques below. \cite{moham2018} conducted a simulation study to investigate and compare the perfomances of the three estimation methods and concluded the maximum likelihood estimators are preferred.

Conditional least squares estimators are found by minimizing

$$\sum^n_{t = 2}(X_t - \alpha X_{t - 1} - \mu(1 - \alpha))^2$$,
where $\mu = E(X_t) = \theta + 2/\theta(\theta + 1)$.  Thus, we have the estimates:

\begin{align*}
    \hat{\alpha}_{CLS} &= \frac{(n - 1)\sum^n_{t = 2}(X_{t - 1}X_t) - \sum^n_{t = 2} X_t \sum^n_{t = 2} X_{t - 1}}{(n - 1)\sum^n_{t = 2} X^2_{t - 1} - \left(\sum_{t = 2}^n X_{t - 1}\right)^2} 
    \nonumber \\\\
    \hat{\mu}_{CLS} &= \frac{\sum^n_{t = 2} X_t - \hat{\alpha}_{CLS} \sum^n_{t = 2}  X_{t - 1}}{(n - 1)(1 - \hat{\alpha}_{CLS})} 
        \nonumber \\\\
    \hat{\theta}_{CLS} &= \frac{-(\hat{\mu}_{CLS} - 1) + \sqrt{(\hat{\mu}_{CLS} - 1)^2 + 8\hat{\mu}_{CLS}}}{2\hat{\mu}_{CLS}} 
\end{align*}

The Yule-Walker estimators of $\alpha$ and $\mu$ are
\begin{align*}
    \hat{\alpha}_{YW} &= \frac{\hat{\gamma}_1}{\hat{\gamma}_0} = \frac{\sum^n_{t = 2}(X_t - \bar{X})(X_{t - 1} - \bar{X})}{\sum_{t = 1}^n \left(X_{t} - \bar{X}\right)^2} 
    \nonumber \\\\
    \hat{\mu}_{YW} &= \bar{X} = \frac{\hat{\theta} + 2}{\hat{\theta}(\hat{\theta} + 1)}
        \nonumber \\\\
    \hat{\theta}_{YW} &= \frac{-(\bar{X} - 1) + \sqrt{(\bar{X} - 1)^2 + 8\bar{X}}}{2\bar{X}} 
\end{align*}
\cite{moham2018} prove that the asymptotic normal distributions for both the conditional least squares estimators and the Yule-Walker estimators are the same.

Finally, the maximum likelihood estimators of $\alpha$ and $\theta$ are the values $\hat\alpha_{ML}$ and $\hat\theta_{ML}$ that maximize the log-likelihood given by 
\begin{align*}
    \ln L(X_1, \hdots, X_n; \theta, \alpha) = 2 \log (\theta) - (X_1 + 3) \log (1 + \theta) + \log (2 + \theta + X_1) + \sum_{t = 2}^n \ln p(X_t | X_{t -1}; \theta, \alpha)
\end{align*}

\subsection{Descriptive Measures of Forecasting Accuracy}
\label{sec:accuracy}
To investigate the accuracy of the forecast in Section~\ref{sec:applications}, we partition an observed dataset of length $n$ into two sets.  The longer set, called the training set, contains the first $m$ observations, and is used to estimate the parameters of the models.  The remaining $n - m$ observations are used for evaluating forecasts using each of the following three criteria.  Our design is similar to~\cite{mandb2016}.  They consider a variety of models, but do not match moments, as does our marginal and innovation methods.  Their comparisons only consider PRMSE and PMAD, but not PTP, which we feel provides a clearer picture of which models perform better.  In their simulated studies, they conclude that the discretized Gaussian forecasts are comparable to or out-perform some discrete models.  In their work, as in ours,  the idea of coherency is central.  A coherent forecast is a forecast that is in the support of the process.  When the support is restricted, as is the case with integer autoregressive processes, the conditional mean may produce forecasts that are not coherent.  In that case, the noncoherent value will be rounded in the traditional manner to produce a coherent forecast.  

The first of the three criteria is the $k$-step-ahead prediction root mean square error, PRMSE($k$), defined as
\begin{align*}
    {\text{PRMSE}}(k) &= \sqrt{\frac{1}{n-m-k+1} \sum_{t = m + k}^n \left(X_t - \hat{X}^{me}_t \right)^2},
\end{align*}
where $\hat{X}^{me}_{t + k}$ is estimated $k$-step-ahead conditional mean given in~\eqref{eq:condmean}.

The second forecasting criteria we consider is the $k$-step-ahead prediction mean absolute deviation, PMAD($k$), given by
\begin{align*}
    {\text{PMAD}}(k) &= \frac{1}{n - m - k + 1} \sum_{t= m + k}^n \left| X_t - \hat{X}^{med}_t \right|, 
\end{align*}
where $\hat{X}^{med}_{t + k}$ is the median of the estimated $k$-step-ahead conditional distribution in~\eqref{eq:kstepPMF}.

The third and final criteria used to investigate the performances of the forecasts is the $k$-step-ahead percentage of true prediction, PTP($k$), given by
\begin{align*}
\label{eq:PTP(k)}
    {\text{PTP}}(k) &= \frac{\sum^{n}_{t = m + k} I \left(X_t = \hat{X}_t \right)}{n - m - k + 1} \times 100\%,
\end{align*}
where $I(\cdot)$ is an indicator function and $\hat{X}_t$ is one of the $k$-step-ahead conditional mean rounded to the nearest integer, the median, or the mode of the estimated $k$-step-ahead conditional distribution given in~\eqref{eq:kstepPMF}.

\section{Application to Discrete Valued Time Series Data}
\label{sec:applications}
In this section we consider the data given in Table~\ref{tab:sex_offences}.  Each record is the number of sex offences reported in the 21st police car beat in Pittsburgh for one month, beginning in January 1990 through December 2001 \citep{RBN2009}.  The data is publicly available in the \emph{R} library {\tt tsinteger} under the dataset name {\tt sexoffences}.   In what follows,  we provide a description of the data's general features.  Following that, we determine an optimal autoregressive order using Akaike information criteria (AIC), corrected AIC (AICc) and the Bayesian information criteria (BIC).  We then fit both a PLINAR and Gaussian AR to the series using the suggested optimal AR order.  The models are then used to forecast the data using the methods described in Section~\ref{sec:comparisons}.  Because the parameters of the models must be estimated, forecasts are also compared for three estimation techniques; namely, conditional least squares, Yule-Walker, and maximum likelihood.

\subsection{Data Description}
The 144 observations in Table ~\ref{tab:sex_offences} are counts of sex offences reported.  The data has a high percentage (62.5\%) of observations equal to zero.  The values in the data range from zero to six.  
\begin{table}[!htbp]
    \centering
    \begin{tabular}{ccccccccccccc}\hline
 & Jan & Feb & Mar & Apr & May & Jun & Jul & Aug & Sep & Oct & Nov & Dec \\\hline
1990 & 0 & 0 & 1 & 0 & 0 & 0 & 1 & 0 & 0 & 0 & 1 & 0 \\
1991 & 0 & 0 & 0 & 0 & 0 & 1 & 1 & 0 & 0 & 0 & 1 & 0 \\
1992 & 0 & 0 & 0 & 0 & 1 & 1 & 2 & 1 & 0 & 1 & 0 & 0 \\
1993 & 1 & 2 & 0 & 0 & 0 & 0 & 1 & 0 & 2 & 0 & 0 & 0 \\
1994 & 0 & 0 & 0 & 2 & 0 & 2 & 0 & 1 & 0 & 3 & 1 & 0 \\
1995 & 1 & 1 & 1 & 0 & 3 & 1 & 0 & 0 & 1 & 2 & 2 & 0 \\
1996 & 0 & 0 & 0 & 0 & 0 & 1 & 1 & 0 & 0 & 0 & 0 & 0 \\
1997 & 0 & 0 & 0 & 1 & 0 & 0 & 0 & 0 & 1 & 0 & 0 & 0 \\
1998 & 0 & 0 & 0 & 0 & 0 & 1 & 2 & 2 & 0 & 2 & 0 & 0 \\
1999 & 1 & 1 & 0 & 3 & 2 & 0 & 0 & 2 & 0 & 0 & 0 & 0 \\
2000 & 1 & 1 & 6 & 5 & 1 & 1 & 0 & 1 & 0 & 0 & 1 & 0 \\
2001 & 0 & 1 & 1 & 0 & 1 & 0 & 1 & 5 & 0 & 0 & 0 & 0 \\\hline
\end{tabular}
    \caption{Number of sex offences reported in the 21st police car beat in Pittsburgh monthly from January 1990 to December 2001.}
    \label{tab:sex_offences}
\end{table}

\begin{figure}[!htbp]
    \centering
   \includegraphics[scale = 0.6]{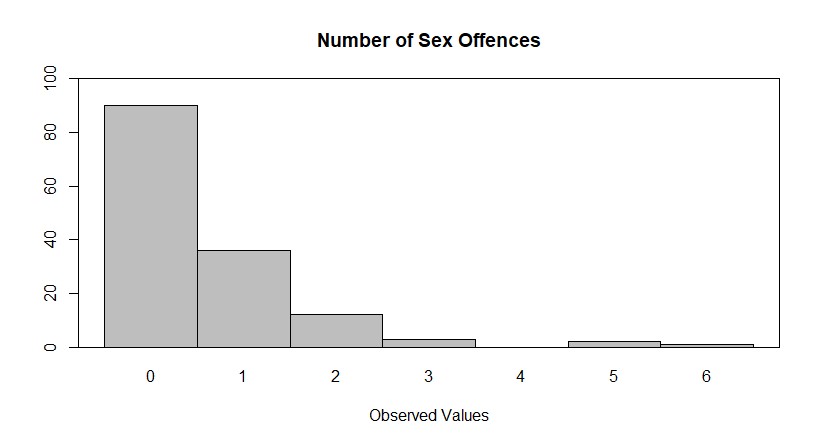}
   \caption{Frequency histogram of the number of sex offences reported in the 21st police car beat in Pittsburgh monthly from January 1990 to December 2001.}
   \label{fig:sex_bar}
\end{figure}

\begin{table}[!htbp]
    \centering
    \begin{tabular}{cccccc}\hline
Minimum & Maximum & Median & Mode & Mean & Variance \\\hline
0 & 6 & 0 & 0 & 0.5903 & 1.0268 \\\hline
\end{tabular}
    \caption{Summary statistics for number of sex offences reported in the 21st police car beat in Pittsburgh monthly from January 1990 to December 2001.}
    \label{tab:sex_offences_summary_stats}
\end{table}

The histogram in Figure~\ref{fig:sex_bar} reveals a right-skewed distribution. The value of zero is the mode of the distribution.  For the 144 observed records, 90 (62.5\%) have a value of zero; 36 (25.00\%) have a value of one; 12 (8.33\%) a value of two, and 3 (2.08\%) a value of three.  No records have a value of four, and only three (2.08\%) have a value of five or six.  Table~\ref{tab:sex_offences_summary_stats} contains the summary statistics for the data.   The median of the data is zero.  The mean number of sex offences is $\bar X = 0.5903$, and the variance is $s_X^2 = 1.0268$.  The index of dispersion is $\hat I = s^2_X/\bar X = 1.7395$, which is large enough to conclude that the data is over-dispersed, which provides a good reason for considering Poisson-Lindley marginals. 

\begin{figure}[!htbp]
    \centering
   \includegraphics[scale = 0.4]{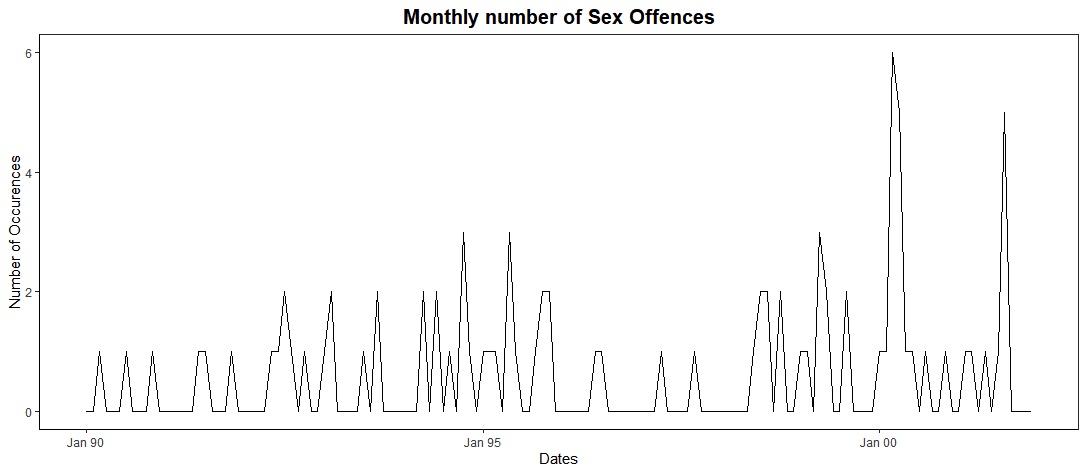}
   \caption{Time plot of number of sex offences reported in the 21st police car beat in Pittsburgh monthly from January 1990 to December 2001. }
   \label{fig:sex_plot}
   \includegraphics[scale = 0.4]{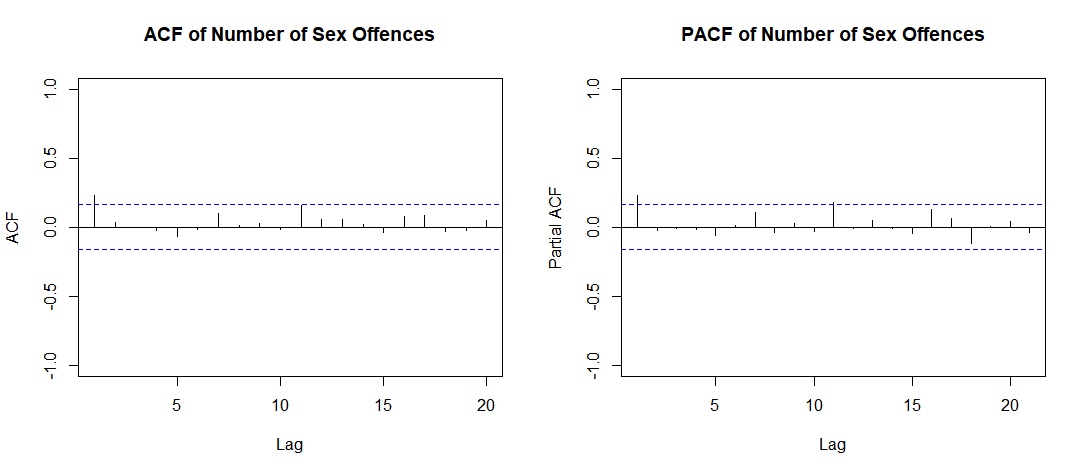}
   \label{fig:sex_corr}
  \caption{Sample autocorrelation function (left) and sample partial autocorrelation function (right) for the sex offences data with large sample 95\% confidence bounds for white noise superimposed (blue dashed lines).}
\end{figure}

Figure~\ref{fig:sex_plot} contains the time plot of the number of sex offences reported in the 21st police car beat in Pittsburgh monthly from January 1990 to December 2001. The data record (Table~\ref{tab:sex_offences}) and the time plot both show that the number of sex offences in the winter months (December - March) is much less than during the remainder of the year. The month-to-month values tend to be close to the same.  In March of 2000 there were six sex offences and in April of 2000 and August of 2001 there were five, both of which are unusually high values.   Figure~\ref{fig:sex_corr} contains the sample autocorrelation function (left) and sample partial autocorrelation function (right) with large sample 95\% confidence bounds for white noise superimposed.  The sample ACF indicates lag one is significant.  The sample PACF has only lag one as significant as well. The indication is that an AR(1) would be a good fit to the data. 
To further aid in determining the optimal AR order for fitting a model, we compute AIC, AICc, and BIC.  Those values are supplied in Table~\ref{tab:sex_offences_AR_criteria}.  All three measures indicate that the optimal order for the AR model is one. 
\begin{table}[!htbp]
    \centering
    \begin{tabular}{cccc}\hline
& AIC & AICc & BIC \\\hline
AR(1) & \textbf{409.3019} & \textbf{409.4734} & \textbf{418.2114} \\\hline
AR(2) & 411.2420 & 411.5298 & 423.1212 \\\hline
AR(3) & 413.2324 & 413.6671 & 428.0814 \\\hline
\end{tabular}
    \caption{Autoregressive order comparison for number of sex offences reported in the 21st police car beat in Pittsburgh monthly from January 1990 to December 2001.}
    \label{tab:sex_offences_AR_criteria}
\end{table}

\clearpage

Table~\ref{tab:sex_offences_estimates} provides estimates for $\alpha$ and $\theta$ using the conditional least squares (CLS), Yule-Walker (YW) and maximum likelihood (ML) when fitting the PLINAR(1) model.  We note that the ML estimates for $\alpha$ is smaller than the CLS and YW estimates.  The small value of the MLE can be attributed to the large number of zeros in the data.
\begin{table}[!htbp]
\centering
\begin{tabular}{c|ccc} \hline
         &   CLS  &   YW   &   ML   \\ \cline{2-4}
$\alpha$ & 0.2297 & 0.2291 & 0.1028 \\
$\theta$ & 2.1671 & 2.1804 & 2.1900 \\ \hline
\end{tabular}
\caption{Conditional least squares, Yule-Walker, and maximum likelihood estimates of $\alpha$ and $\theta$ for the sex offences data.}
\label{tab:sex_offences_estimates}
\end{table}

Tables~\ref{tab:sex_offences_plinar_cpmf},~\ref{tab:sex_offences_marginal_cpmf}, and~\ref{tab:sex_offences_innov_cpmf} provide the conditional $k$-step-ahead probability mass functions given $X_n = 0$ where model parameters are estimated using each of CLS, YW, and ML based on the traditional PLINAR(1) model, the model using the marginal means method, and the model using the innovation method, respectively.  
To find the estimates of the PMFs, the first 141 values of the data were used.  In this study, we consider only $X_n = 0$ since all four of observation numbers 141 through 144 are zero.  In subsequent application, conditional $k$-step-ahead forecasts for values of $X_n$ other than zero are considered.  The conditional PMFs are used for computing the means, medians, and modes that are used for the $k$-step-ahead forecasts in Table~\ref{tab:sex_offences_forecasts} for the PLINAR(1), marginal, and innovation methods, respectively. As well as the traditional Gaussian AR(1) was considered. We note that Table~\ref{tab:sex_offences_plinar_cpmf} provides the conditional PMFs for only up to $X_{n + k} = 8$, and using these values one is able to compute the means found in Table~\ref{tab:sex_offences_forecasts}.

\begin{sidewaystable}[!htbp]
    \centering
    \begin{tabular}{c|cccc|cccc|cccc}
    \hline
& \multicolumn{4}{c|}{CLS} & \multicolumn{4}{c|}{YW} & \multicolumn{4}{c}{MLE} \\ \cline{1-13}
& $k = 1$ & $k = 2$ & $k = 3$ & $k = \infty$ & $k = 1$ & $k = 2$ & $k = 3$ & $k = \infty$ & $k = 1$ & $k = 2$ & $k = 3$ & $k = \infty$ \\   \hline
 $p_k(0|0)$ & 0.703 & 0.636 & 0.621 & 0.616 & 0.704 & 0.637 & 0.622 & 0.618 & 0.657 & 0.623 & 0.619 & 0.619 \\ 
  $p_k(1|0)$ & 0.188 & 0.229 & 0.238 & 0.241 & 0.188 & 0.229 & 0.238 & 0.241 & 0.217 & 0.238 & 0.240 & 0.240 \\ 
  $p_k(2|0)$ & 0.070 & 0.086 & 0.090 & 0.091 & 0.070 & 0.086 & 0.089 & 0.090 & 0.081 & 0.089 & 0.090 & 0.090 \\ 
  $p_k(3|0)$ & 0.025 & 0.032 & 0.033 & 0.033 & 0.025 & 0.031 & 0.033 & 0.033 & 0.029 & 0.032 & 0.033 & 0.033 \\ 
  $p_k(4|0)$ & 0.009 & 0.011 & 0.012 & 0.012 & 0.009 & 0.011 & 0.012 & 0.012 & 0.010 & 0.012 & 0.012 & 0.012 \\ 
  $p_k(5|0)$ & 0.003 & 0.004 & 0.004 & 0.004 & 0.003 & 0.004 & 0.004 & 0.004 & 0.004 & 0.004 & 0.004 & 0.004 \\ 
  $p_k(6|0)$ & 0.001 & 0.001 & 0.001 & 0.001 & 0.001 & 0.001 & 0.001 & 0.001 & 0.001 & 0.001 & 0.001 & 0.001 \\ 
  $p_k(7|0)$ & 0.000 & 0.000 & 0.001 & 0.001 & 0.000 & 0.000 & 0.000 & 0.001 & 0.000 & 0.000 & 0.000 & 0.000 \\ 
  $p_k(8|0)$ & 0.000 & 0.000 & 0.000 & 0.000 & 0.000 & 0.000 & 0.000 & 0.000 & 0.000 & 0.000 & 0.000 & 0.000 \\ 
   \hline
\end{tabular}
\caption{Estimated conditional $k$-step-ahead probability mass function for $X_n = 0$ for three estimation methods for $\alpha$ and $\theta$ based on PLINAR(1) model for the sex offences data.}
    \label{tab:sex_offences_plinar_cpmf}
\end{sidewaystable}
\clearpage

\begin{sidewaystable}[!htbp]
    \centering
    \begin{tabular}{c|cccc|cccc|cccc}
    \hline
& \multicolumn{4}{c|}{CLS} & \multicolumn{4}{c|}{YW} & \multicolumn{4}{c}{ML} \\ \cline{1-13}
& $k = 1$ & $k = 2$ & $k = 3$ & $k = \infty$ & $k = 1$ & $k = 2$ & $k = 3$ & $k = \infty$ & $k = 1$ & $k = 2$ & $k = 3$ & $k = \infty$ \\     \hline
$p_k(0|0)$ & 0.309 & 0.276 & 0.267 & 0.265 & 0.310 & 0.276 & 0.268 & 0.265 & 0.286 & 0.268 & 0.266 & 0.266 \\ 
  $p_k(1|0)$ & 0.405 & 0.395 & 0.393 & 0.393 & 0.406 & 0.396 & 0.395 & 0.395 & 0.400 & 0.396 & 0.396 & 0.396 \\ 
  $p_k(2|0)$ & 0.234 & 0.260 & 0.266 & 0.267 & 0.233 & 0.259 & 0.265 & 0.267 & 0.251 & 0.265 & 0.266 & 0.266 \\ 
  $p_k(3|0)$ & 0.048 & 0.064 & 0.067 & 0.068 & 0.047 & 0.063 & 0.066 & 0.067 & 0.058 & 0.065 & 0.066 & 0.066 \\ 
  $p_k(4|0)$ & 0.004 & 0.006 & 0.006 & 0.006 & 0.003 & 0.005 & 0.006 & 0.006 & 0.005 & 0.006 & 0.006 & 0.006 \\ 
  $p_k(5|0)$ & 0.000 & 0.000 & 0.000 & 0.000 & 0.000 & 0.000 & 0.000 & 0.000 & 0.000 & 0.000 & 0.000 & 0.000 \\ 
  $p_k(6|0)$ & 0.000 & 0.000 & 0.000 & 0.000 & 0.000 & 0.000 & 0.000 & 0.000 & 0.000 & 0.000 & 0.000 & 0.000 \\ 
  $p_k(7|0)$ & 0.000 & 0.000 & 0.000 & 0.000 & 0.000 & 0.000 & 0.000 & 0.000 & 0.000 & 0.000 & 0.000 & 0.000 \\ 
  $p_k(8|0)$ & 0.000 & 0.000 & 0.000 & 0.000 & 0.000 & 0.000 & 0.000 & 0.000 & 0.000 & 0.000 & 0.000 & 0.000 \\  
   \hline
\end{tabular}
\caption{Estimated conditional $k$-step-ahead probability mass function fo $X_n = 0$ for three estimation methods for $\alpha$ and $\theta$ based on marginal method for the sex offences data.}
    \label{tab:sex_offences_marginal_cpmf}
\end{sidewaystable}

\begin{sidewaystable}[!htbp]
    \centering
    \begin{tabular}{c|cccc|cccc|cccc}
    \hline
& \multicolumn{4}{c|}{CLS} & \multicolumn{4}{c|}{YW} & \multicolumn{4}{c}{ML} \\ \cline{1-13}
& $k = 1$ & $k = 2$ & $k = 3$ & $k = \infty$ & $k = 1$ & $k = 2$ & $k = 3$ & $k = \infty$ & $k = 1$ & $k = 2$ & $k = 3$ & $k = \infty$ \\     \hline
$p_k(0|0)$ & 0.298 & 0.262 & 0.254 & 0.251 & 0.298 & 0.263 & 0.254 & 0.252 & 0.280 & 0.261 & 0.259 & 0.259 \\ 
  $p_k(1|0)$ & 0.429 & 0.418 & 0.417 & 0.416 & 0.431 & 0.420 & 0.419 & 0.418 & 0.411 & 0.408 & 0.408 & 0.407 \\ 
  $p_k(2|0)$ & 0.232 & 0.262 & 0.268 & 0.270 & 0.231 & 0.261 & 0.267 & 0.269 & 0.252 & 0.266 & 0.267 & 0.267 \\ 
  $p_k(3|0)$ & 0.039 & 0.054 & 0.057 & 0.058 & 0.038 & 0.053 & 0.056 & 0.057 & 0.053 & 0.060 & 0.061 & 0.061 \\ 
  $p_k(4|0)$ & 0.002 & 0.004 & 0.004 & 0.004 & 0.002 & 0.003 & 0.004 & 0.004 & 0.004 & 0.005 & 0.005 & 0.005 \\ 
  $p_k(5|0)$ & 0.000 & 0.000 & 0.000 & 0.000 & 0.000 & 0.000 & 0.000 & 0.000 & 0.000 & 0.000 & 0.000 & 0.000 \\ 
  $p_k(6|0)$ & 0.000 & 0.000 & 0.000 & 0.000 & 0.000 & 0.000 & 0.000 & 0.000 & 0.000 & 0.000 & 0.000 & 0.000 \\ 
  $p_k(7|0)$ & 0.000 & 0.000 & 0.000 & 0.000 & 0.000 & 0.000 & 0.000 & 0.000 & 0.000 & 0.000 & 0.000 & 0.000 \\ 
  $p_k(8|0)$ & 0.000 & 0.000 & 0.000 & 0.000 & 0.000 & 0.000 & 0.000 & 0.000 & 0.000 & 0.000 & 0.000 & 0.000 \\ 
   \hline
\end{tabular}
\caption{Estimated conditional $k$-step-ahead probability mass function fo $X_n = 0$ for three estimation methods for $\alpha$ and $\theta$ based on innovation method for the sex offences data.}
    \label{tab:sex_offences_innov_cpmf}
\end{sidewaystable}
\clearpage

\begin{table}[!htbp]
    \centering
    \begin{tabular}{c|c|ccc|ccc|ccc}
    \hline
     & & \multicolumn{3}{c|}{CLS} & \multicolumn{3}{c|}{YW} & \multicolumn{3}{c}{ML} \\ \cline{1-11}
& & $k = 1$ & $k = 2$ & $k = 3$ & $k = 1$ & $k = 2$ & $k = 3$ & $k = 1$ & $k = 2$ & $k = 3$  \\     \hline
\multirow{3}{*}{\rotatebox[origin = c]{90}{\parbox[c]{18mm}{\centering PLINAR(1)}}} & Mean & 0.468 & 0.575 & 0.600 & 0.465 & 0.571 & 0.596 & 0.538 & 0.593 & 0.599 \\ [2mm]
  & Med. & 0.000 & 0.000 & 0.000 & 0.000 & 0.000 & 0.000 & 0.000 & 0.000 & 0.000 \\ [2mm]
  & Mode & 0.000 & 0.000 & 0.000 & 0.000 & 0.000 & 0.000 & 0.000 & 0.000 & 0.000 \\ [2mm]
  \hline
\multirow{3}{*}{\rotatebox[origin = c]{90}{\parbox[c] {18mm}{\centering Marginal Method}}} &   Mean & 0.468 & 0.575 & 0.600 & 0.465 & 0.571 & 0.596 & 0.538 & 0.593 & 0.599 \\ [2mm]
 & Med. & 1.000 & 1.000 & 1.000 & 1.000 & 1.000 & 1.000 & 1.000 & 1.000 & 1.000 \\ [2mm]
  & Mode & 1.000 & 1.000 & 1.000 & 1.000 & 1.000 & 1.000 & 1.000 & 1.000 & 1.000 \\ [2mm]
   \hline
   \multirow{3}{*}{\rotatebox[origin = c]{90}{\parbox[c] {18mm}{\centering Innovation Method}}} &   Mean & 0.468 & 0.575 & 0.600 & 0.465 & 0.571 & 0.596 & 0.538 & 0.593 & 0.599 \\ [2mm]
 & Med. & 1.000 & 1.000 & 1.000 & 1.000 & 1.000 & 1.000 & 1.000 & 1.000 & 1.000 \\ [2mm]
 & Mode & 1.000 & 1.000 & 1.000 & 1.000 & 1.000 & 1.000 & 1.000 & 1.000 & 1.000 \\ [2mm]
   \hline
\multirow{2}{*}{\rotatebox[origin = c]{90}
{\parbox[c]{12mm}{\centering Gauss.}}}    &  &  &  &  &  &  & & & &  \\ 
& AR(1)  & 0.468 & 0.575 & 0.600 & 0.465 & 0.571 & 0.596 & 0.463 & 0.569 & 0.593  \\ [4mm]
   \hline 
& Truth & 0 & 0 & 0 & 0 & 0 & 0 & 0 & 0 & 0 \\ 
\hline
\end{tabular}
\caption{Conditional $k$-step-ahead forecasts for the sex offences data for each of the PLINAR(1) model, the marginal method, innovation method.  For each of the three discrete model methods, model parameters were estimated using the three estimation methods in Section~\ref{sec:estimation}.  For the Gaussian AR(1) and Gaussian AR(2), model parameters were estimated using ordinary least squares, Yule-Walker, or maximum likelihood via the \emph{R} function \tt ar.ols, ar.yw, and ar.mle.}
    \label{tab:sex_offences_forecasts}
\end{table}

For this data $X_{141} = X_{142} = X_{143} = X_{144} = 0$. From Table~\ref{tab:sex_offences_forecasts}, the median and mode of the PLINAR(1) all provide correct forecasts for $k = 1, 2$ and 3. The rounded mean for all four models provides a correct forecast for one-step-ahead.   However, for $k = 2$ and 3, the means incorrectly forecasts a value of one.  For the marginal and innovation methods, the median and modes incorrectly forecast a value of one for all $k = 1, 2$, and 3.  

\clearpage

\begin{sidewaystable}[!htbp]
    \centering
    \begin{tabular}{c|c|ccc|ccc|ccc|ccc}
    \hline
    & & \multicolumn{3}{c|}{PRMSE} & \multicolumn{3}{c|}{PMAD} & \multicolumn{3}{c|}{PTP (Med.)} & \multicolumn{3}{c}{PTP (Mode)}\\ \cline{1-14}
 & & $k = 1$ & $k = 2$ & $k = 3$  & $k = 1$ & $k = 2$ & $k = 3$ & $k = 1$ & $k = 2$ & $k = 3$  & $k = 1$ & $k = 2$ & $k = 3$ \\     \hline
\multirow{4}{*}{\rotatebox[origin = c]{90}{\parbox[c]{18mm}{\centering CLS}}} & PLINAR(1) & 1.597 & 1.669 & 1.700 & 0.931 & 0.929 & 0.963 & 0.517 & 0.536 & 0.519 & 0.517 & 0.536 & 0.519 \\ 
  & Marginal & 1.597 & 1.669 & 1.700 & 0.966 & 1.000 & 1.000 & 0.345 & 0.357 & 0.370 & 0.345 & 0.357 & 0.370 \\ 
  & Innovation & 1.597 & 1.669 & 1.700 & 0.966 & 1.000 & 1.000 & 0.345 & 0.357 & 0.370 & 0.345 & 0.357 & 0.370 \\ 
  & Traditional & 1.597 & 1.669 & 1.700 & 0.966 & 0.929 & 0.889 & 0.414 & 0.464 & 0.519 & 0.414 & 0.464 & 0.519 \\
   \hline
\multirow{4}{*}{\rotatebox[origin = c]{90}{\parbox[c] {18mm}{\centering YW}}} & PLINAR(1) & 1.597 & 1.669 & 1.895 & 0.931 & 0.929 & 0.963 & 0.517 & 0.536 & 0.519 & 0.517 & 0.536 & 0.519 \\ 
  & Marginal & 1.597 & 1.669 & 1.895 & 0.966 & 1.000 & 1.000 & 0.345 & 0.357 & 0.370 & 0.345 & 0.357 & 0.370 \\ 
  & Innovation & 1.597 & 1.669 & 1.895 & 0.966 & 1.000 & 1.000 & 0.345 & 0.357 & 0.370 & 0.345 & 0.357 & 0.370 \\ 
  *& Traditional & 1.597 & 1.669 & 1.895 & 0.966 & 0.929 & 1.000 & 0.414 & 0.464 & 0.481 & 0.414 & 0.464 & 0.481 \\ 
   \hline
   \multirow{4}{*}{\rotatebox[origin = c]{90}{\parbox[c] {18mm}{\centering ML }}} & PLINAR(1) & 1.597 & 1.852 & 1.886 & 0.931 & 0.929 & 0.963 & 0.517 & 0.536 & 0.519 & 0.517 & 0.536 & 0.519 \\ 
  & Marginal & 1.597 & 1.852 & 1.886 & 0.966 & 1.000 & 1.000 & 0.345 & 0.357 & 0.370 & 0.345 & 0.357 & 0.370 \\ 
  & Innovation & 1.597 & 1.852 & 1.886 & 0.966 & 1.000 & 1.000 & 0.345 & 0.357 & 0.370 & 0.345 & 0.357 & 0.370 \\ 
  & Traditional & 1.597 & 1.669 & 1.895 & 0.966 & 0.929 & 1.000 & 0.414 & 0.464 & 0.481 & 0.414 & 0.464 & 0.481 \\ 
   \hline
\end{tabular}
\caption{Accuracy measures of forecasts of sex offences data based on a PLINAR(1) model, the marginal method, innovation method, and traditional Gaussian AR(1) where model parameters are estimated using conditional least squares (CLS), Yule-Walker (YW) and maximum likelihood (ML).}
    \label{tab:sex_offences_accuracy}
\end{sidewaystable}

The measures of forecasts accuracy in Table~\ref{tab:sex_offences_accuracy} were computed training the models on the first 80\% (115 observations) of the data.  From that training set, $k$-step-ahead forecasts were computed for the remaining 20\% (29 observations) for $k = 1, 2$, and 3.  Focusing on PRMSE and PMAD, the forecasts using CLS, YW and ML estimates for the models perform equally well. Focusing on the PTP, the median and mode performed equally well for all four processes.

\section{Conclusions}
\label{sec:conclusion}
In this paper, we investigated forecasting discrete time series using a PLINAR(1) model, a traditional Gaussian AR(1) model, and two modified version of the traditional Gaussian AR(1) model.  The resultant distribution functions were compared via Kullback-Liebler divergence and Kolmogorov metric for a variety of values of $\alpha$ and $\theta$.  To determine which of the marginal method or innovation method performs best in approximating the conditional Poisson-Lindley PMF, the innovation method would be preferred when it is believed that values of $\alpha$ and $\theta$ are close to zero or one, since the innovation method has a smaller standard error compared to the marginal method, and the two approximations perform equally well.  However, for moderate values of $\alpha$ and moderate to large values of $\theta$, the distances between the marginal approximation and the conditional Poisson-Lindley PMF are smaller, and so the marginal method would be preferred.

One application was considered to further evaluate and compare the forecasting potential of the PLINAR(1), and the marginal, innovation, and traditional Gaussian AR for discrete-valued time series.  This data set had overdispersion, which could provide justification for using the PLINAR.  For this data set, the PLINAR(1) model outperformed its Gaussian counterparts.  It should also be noted that the zero-inflated Poisson distribution may be a better choice of distribution for modeling this data, and that work is being considered in additional research.

The authors would like to thank Annika Homburg for helpful discussions and allowing us to use her code based on the Poisson distribution to help develop our own for the Poisson-Lindley.  
\bibliographystyle{apalike}
\bibliography{references}

\newpage 

\end{document}